\documentclass[10pt,twocolumn]{IEEEtran} 
\usepackage{times}
\usepackage{float}
\usepackage{subfigure}
\usepackage{float}
\usepackage{amsmath}
\usepackage{color}
\usepackage{graphicx}
\usepackage{epsfig}

\usepackage{algorithm}
\usepackage[noend]{algorithmic}

\begin{document}
\title{Inferring Neuronal Network Connectivity using Time-constrained Episodes}

\author{
Debprakash Patnaik
(Electrical Engg. Dept.,
Indian Institute of Science,
Bangalore), and \\
P. S. Sastry
(Electrical Engg. Dept.,
Indian Institute of Science,
Bangalore), and\\
K. P. Unnikrishnan
(General Motors R\&D,
Warren)
}

\maketitle
\thispagestyle{empty}

\begin{abstract}
Discovering frequent episodes in event sequences is an 
interesting data mining task. In this paper, we argue that this framework is very effective 
for analyzing multi-neuronal spike train data. Analyzing spike train 
data is an important problem in neuroscience though there are no data mining 
approaches reported for this. 
Motivated by this application, we introduce different temporal constraints on 
the occurrences of episodes.  
We present algorithms for discovering frequent episodes under temporal 
constraints. Through simulations, we show that our method is very effective for 
analyzing spike train data for unearthing underlying connectivity patterns. 

\end{abstract}

\section{Introduction}

Temporal data mining is concerned with mining of large sequential data
sets \cite{Srivats-survey2005}. Frequent episode discovery, originally proposed 
in \cite{Mannila1997}, is one of the popular frameworks in temporal data mining. 
Here the data is viewed as a single long sequence of events and the task is to 
unearth temporal patterns (called episodes) that occur sufficiently often along 
that sequence. Examples of such data are alarms in a telecommunication network, 
fault logs of a manufacturing plant, multi-neuronal spike train recordings,
etc. 

In this paper we present new algorithms for discovering frequent episodes 
under some temporal constraints. The motivation for considering 
such constraints comes from the application that we discuss here, 
namely, analyzing multi-neuron spiking data to infer useful information 
about the underlying microcircuits. Such neuron spike train data 
can be obtained through 
techniques such as microelectrode array experiments. Analyzing simultaneously 
recorded data from a number of neurons is an important and challenging problem.
The data consists of spike trains from a number of neurons. Since functionally 
interconnected neurons tend to fire in certain precise patterns, discovering 
frequent patterns in such temporal data can help understand the underlying 
neural circuitry. Here, we argue that the frequent episodes framework is ideally 
suited for such analysis. 
However, as we shall see, for this application we 
need methods to discover frequent episodes where the occurrences of 
episodes need to satisfy some additional temporal constraints. 
The currently available methods for frequent episode discovery can not 
tackle such constraints. In this paper, we present some new algorithms for 
frequent episode discovery under such temporal constraints. 

We explain the problem of analyzing multi-neuronal spike train data in 
Section~\ref{sec:mea}. We then present a 
brief overview of the frequent episodes framework in Section~\ref{sec:epi}. 
We introduce the notion of temporal constraints on the episode 
occurrences and explain how one can use methods of serial and parallel episode 
discovery under temporal constraints, to discover many patterns of interest in the 
spike train data. The algorithms for discovering frequent episodes under temporal 
constraints are presented in Section~\ref{sec:algos}. We present some simulation 
results to illustrate our method of discovering connection patterns in neuronal 
networks in Section~\ref{sec:res}. Finally, we conclude the paper in 
Section~\ref{sec:conc} with a discussion. 

\section{Multi-neuronal Spike Train Data}
\label{sec:mea}

Over the last couple of decades many new technologies have made it 
possible to simultaneously record signals from many neurons and 
hence to study microcircuits in neuronal assemblies. 
Microelectrode array (MEA) is one such popular recording 
technology. A typical MEA setup consists of $8\times 8$ grid of 64 electrodes 
with inter-electrode spacing of about 25 microns and can be mounted on a 
neural culture or brain slice. Other technologies for recording from multiple 
neurons include imaging of neuronal currents using some specialized dyes.  These 
technologies now allow for gathering of vast amounts of data, especially in 
neuronal cultures, using which one wishes to study connectivity patterns and 
microcircuits in neural systems. (See \cite{Potter2006,Ikegaya2004} for some 
recent studies of this kind).


The availability of vast amounts of such data means that developing efficient methods 
to analyze neuronal spike trains is a challenging task of immediate utility 
in this area. A recent review by Brown et.al. summarizes the current state of 
art \cite{Brown2004}. 
Most of the current methods of analysis rely on quantities that can be 
computed through cross correlations among spike trains (time shifted with respect to 
one another) to identify interesting patterns in spiking activity \cite{Brown2004}. 
There are also methods that look for specific fixed patterns and assess their 
statistical significance under a null hypothesis that different spike trains are 
\textit{iid} Bernoulli processes \cite{Abeles1988,Lee2004,Tetko2001}. Most such 
methods can not look for patterns that involve more than 3 or 4 neurons due to the 
ubiquitous curse of dimensionality. Hence model-free techniques such as data mining 
can be very useful in unearthing interesting patterns in the 
spike trains. 

The patterns that one is interested in this application can be roughly grouped into 
what are called Synchrony, Order and Synfire chains. Synchronous firing by a group of 
neurons is interesting because it can be an efficient way to transmit information 
\cite{Meister2005}. Ordered firing sequences of neurons where times between 
firing of successive neurons are fairly constant denote a chain of triggering 
events and unearthing such relations between neurons can thus reveal some 
microcircuits \cite{Abeles2001}. 
Such an ordered chain may be among groups of neurons rather than 
single neurons. Such a pattern is called a Synfire chain and is believed to be 
a very important microcircuit \cite{Ikegaya2004}. 
In the next section, we explain how all such 
patterns can be discovered under the framework of frequent episodes with 
temporal constraints.

\section{Frequent Episode Discovery}
\label{sec:epi}

Frequent episode discovery framework was proposed by Mannila et.al. 
\cite{Mannila1997} in the context analyzing alarm sequences in a communication 
network. Laxman et.al. \cite{Srivats2005} introduced the notion of 
non-overlapped occurrences as episode frequency and proposed  
efficient counting algorithms. We first give a brief overview of this framework.

The data to be analyzed is a sequence of events 
denoted by $\langle(E_{1},t_{1}),(E_{2},t_{2}),\ldots\rangle$ where $E_{i}$ 
represents the \textit{event type} and $t_{i}$ the \textit{time of occurrence} of 
the $i^{th}$ event. $E_i$'s are drawn from a finite set of event types. 
The sequence is ordered with respect to times of occurrences so 
that, $t_i\le t_{i+1}$, $\forall i$. The following is an 
example event sequence containing 7 events with 5 event types.
\begin{equation}
\langle(A,1),(B,3),(D,4),(C,6),(A,12),(E,14),(B,15)\rangle
\label{eq:data-seq}
\end{equation}

In multi-neuron data, a spike event has the label of the neuron (or the electrode 
number in case of micro-electrode array recordings) which 
generated the spike as its event type  and has the associated time of occurrence. 
The neurons in the ensemble under observation fire action potentials (or spikes) at 
different times. All these spike events are 
strung together, in time order, to give a single long data sequence as needed for 
frequent episode discovery.

The general temporal patterns that we wish to discover in this 
framework are called episodes. 
 In this paper we shall deal with two types of 
episodes: \textit{Serial} and \textit{Parallel}.

A \textit{serial episode} is an ordered tuple of event types. For example, 
$(A\rightarrow B\rightarrow C)$ is a 3-node serial episode. The arrows in this 
notation indicate the order of the events. Such an episode is said to 
\textit{occur} in an event sequence if there are  corresponding events in the 
prescribed order. In sequence (\ref{eq:data-seq}), the events 
{${(A,1),(B,3),(C,6)}$} constitute an occurrence of the above episode. In 
contrast a \textit{parallel episode} is similar to an unordered set of items. We denote a 
3-node parallel episode with event types $A$, $B$ and $C$, as $(ABC)$. An 
occurrence of $(ABC)$ can have the events in any order in the sequence. The events
{${(B,3),(C,6),(A,12)}$} constitute an occurrence of the parallel episode $(ABC)$.

We note here that occurrence of an episode (of either type) does not 
require the associated event types to occur consecutively;  
there can be other intervening events between them. 
 In the multi-neuronal data, if neuron $A$ makes 
neuron $B$ to fire, then, we expect to see $B$ following $A$ often. However, in 
different occurrences of such a substring, there may be different number of 
other spikes between $A$ and $B$ because many other neurons may also be spiking 
simultaneously. Thus, the episode structure allows us to unearth patterns  
in the presence of such noise in spike data.

An episode $\beta$ is a sub-episode of episode $\alpha$ if all event types of 
$\beta$ are in $\alpha$ and if the order among the event types of $\beta$ is 
same as that for the corresponding event types in $\alpha$. For example 
$(A\rightarrow B)$, $(A\rightarrow C)$, and $(B\rightarrow C)$ are 2-node 
sub-episodes of the 3-node episode $(A\rightarrow B\rightarrow C)$, while 
$(B\rightarrow A)$ is not. In case of parallel episodes, there is no ordering 
requirement. Hence every subset of the set of event types of an episode is a 
subepisode. It is to be noted here that occurrence of an episode implies 
occurrence of all its subepisodes.


Frequency of an episode is some measure of how often an episode occurs in the 
data and there are different ways of defining it.
Here, we use the frequency measure proposed in \cite{Srivats2005} known 
as \textit{non-overlapped occurrence} count. A collection occurrences of an 
episode $\alpha$ are said to be \textit{non-overlapped} if no event associated 
with one appears in between the events associated with any other. 
The corresponding frequency for episode $\alpha$ is 
defined as the cardinality of the largest set of non-overlapped occurrences of 
$\alpha$ in the given event sequence. (See \cite{Srivats2005} for more discussion). 
This definition of frequency results in very efficient 
counting algorithms \cite{Srivats2005}. 
In the context of our application, counting non-overlapped occurrences is 
natural because we would then be looking at causative chains that happen at 
different times again and again.

\subsection{Temporal Constraints}
As stated earlier, in this paper we present algorithms for discovering frequent 
episodes where, while counting the frequency, we include only those occurrences which 
satisfy some additional temporal constraints. We mainly consider two types of 
such constraints: episode expiry time and inter-event time constraints. 

Given an episode occurrence (that is, a set of events in the data stream that 
constitute an occurrence of the episode), we call the largest time difference 
between any two events constituting the occurrence as the span of the occurrence. 
For serial episodes, this would be the difference between times of the first and 
the last events. The episode expiry time constraint 
requires that we count only those occurrences whose span is less than a (user-specified) 
time $T_X$. In the algorithm in \cite{Mannila1997}, the window width essentially 
implements an upper bound on the span of occurrences. An efficient algorithm for 
counting non-overlapping occurrences of serial episodes that satisfy an expiry time 
constraint is available in \cite{Srivats2005}.

The inter-event time constraint, which is meaningful only 
for serial episodes, is specified by giving an interval of  
the form $(T_{low}, T_{high}]$ and requires that the difference between the times of 
 every pair of successive events in any occurrence of a serial episode 
should be in this interval.  In a generalized form of this constraint, we may have 
different time intervals for different pairs of events. 
In the next section, we present algorithms for 
counting non-overlapped occurrences of episodes under time constraints.

While these temporal constraints are motivated by our application, these are 
fairly general and would be useful in many other applications of frequent 
episode discovery. 

\subsection{Episodes as patterns in neuronal spike data}

The analysis requirements of spike train data
are met very well by the frequent episodes framework. Serial
and parallel episodes with appropriate temporal
constraints can capture many patterns of interest in multi-neuronal data. 

\begin{figure}[!htb]
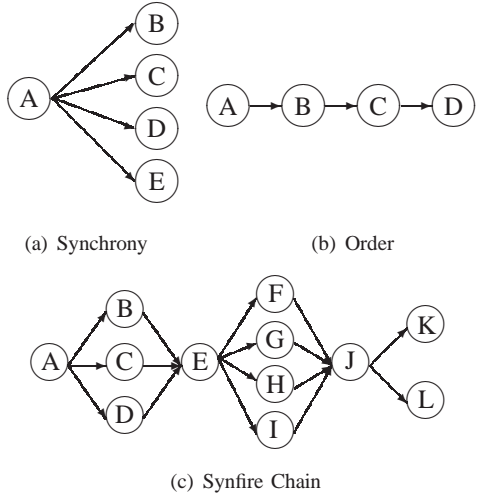

\centering
\subfigure[Synchrony]{\input{parallel-structure.pic}}    
\subfigure[Order]{\input{serial-structure.pic}}
\subfigure[Synfire Chain]{\input{synfire-structure.pic}}
\caption{Examples of neuronal connection structures that can result in different 
patterns in the spike trains}
\label{fig:episode-structures}
\end{figure}

As stated earlier, one of the patterns of interest is
Synchrony or co-spiking activity in which groups of neurons
fire synchronously. This kind of synchrony may not be precise. 
Allowing for some amount of variability, co-spiking
activity requires that all neurons must fire within a small interval
of time of each other (in any order) for them to be grouped together.
Such synchronous firing patterns may be generated using the structure
as shown in Fig.~\ref{fig:episode-structures}(a).
Such patterns of Synchrony can be discovered 
by looking for frequent parallel episodes which satisfy an expiry time constraint.
The expiry time here controls the amount of variability allowed
for declaring a grouped activity as synchronous.

Another pattern in spike data is ordered firings.
A simple mechanism that can generate ordered firing sequences is
shown in Fig.~\ref{fig:episode-structures}(b).
Serial episodes capture such a pattern
very well. Once again, we may need some additional time constraints. 
A useful constraint is that of inter-event time constraint.
In multi-neuron data, if we want to conclude that $A$ is causing $B$ to
fire, then $B$ cannot occur too soon after $A$ because there would be
some propagation delay and $B$ can not occur too much later than $A$
because the effect of firing of $A$ would not last indefinitely.
For example, we can prescribe that inter-event times to be of the same order as the 
synaptic delay times so that a frequent serial episode may capture an underlying 
microcircuit. Thus, serial episodes with proper inter-event time constraints can 
capture ordered firing sequences which may be due to underlying functional 
connectivity.

Another important pattern in spiking data is that of synfire
chains \cite{Ikegaya2004}. This consists of groups of synchronously firing 
neurons strung together with tight temporal constraints. By a combination of 
parallel and serial episode discovery we can unearth such patterns also. The  
structure shown in Fig.~\ref{fig:episode-structures}(c) captures such a synfire 
chain. We can think of this as a microcircuit where $A$ primes synchronous firing 
of $(B C D)$, which, through $E$, causes synchronous firing of $(F G H I)$ and 
so on. When such a pattern occurs often in the spike train data, parallel 
episodes like $(B C D)$ and $(F G H I)$ become frequent (by using appropriate 
expiry time constraint). 
After discovering all such parallel episodes,  we  replace all
recognized occurrences of each of these episodes by a new event in the data 
stream with a new symbol (representing the episode) for the event type and 
an appropriate time of occurrence.  Then we 
discover serial episodes  on this new data stream. With this procedure, 
 we can unearth patterns such as 
synfire chains. We show later that our algorithms can discover such
synfire chains.


\section{Discovering frequent episodes under temporal constraints}
\label{sec:algos}

In this section we describe our algorithms that discover frequent episodes 
under expiry and inter-event time constraints. 
The inter-event time constraints are meaningful only for serial episodes and 
that is the case we consider. 
Since algorithms for 
taking care of expiry time are available in case of serial episodes \cite{Srivats2005}, 
we consider the case of only parallel episodes under expiry time constraint.

 A frequent episode is one whose frequency 
exceeds a user specified threshold. The overall objective is to find all 
frequent episodes. Counting of all possible episodes is infeasible in most real 
problems due to combinatorial explosion. As is common in such data mining methods, 
we use a level-wise Apriori style \cite{Agrawal95} procedure. 
Under this we use frequent $N$-node episodes to create N+1-node candidates, and, 
using another pass over the data, obtain frequent N+1-node episodes. 
The basic structure of the frequency counting algorithm is similar to the ones in 
\cite{Mannila1997,Srivats2005} and we also use 
finite state automata for recognizing episode occurrences. 

\subsection{\label{sec:interval-discovery-algo}Serial Episode with Inter-event 
Constraints}

Under an inter-event time constraint, the time between successive events in any 
occurrence have to be in a prescribed interval. To take care of this we use a 
new episodes structure. The episode structure now consists of an ordered set of 
intervals besides the set of event types. An interval $(t_{low}^i,t_{high}^i]$ is 
associated with $i^{th}$ pair of consecutive of event types in the episode. 
For example, a 4-node serial episode is now denoted as follows:
\begin{equation}
(A^{\underrightarrow{(t_{low}^{1},t_{high}^{1}]}}B^{\underrightarrow{(t_{low}^{2},t_{high}^{2}]}}C^{\underrightarrow{(t_{low}^{3},t_{high}^{3}]}}D)
\label{eq:episode-with-interval}
\end{equation}
In a given occurrence of episode $A\rightarrow$ $B\rightarrow$ $C\rightarrow$ $D$ 
let $t_{A}$, $t_{B}$, $t_{C}$ and $t_{D}$ denote the time of occurrence of corresponding 
event types. Then this is a valid occurrence of the serial episode with inter-event 
time constraint given by (\ref{eq:episode-with-interval}), 
if $t_{low}^{1}$ $<$ $(t_{B}-t_{A})$ $\le$ $t_{high}^{1}$,
$t_{low}^{2}$ $<$ $(t_{C}-t_{B})$ $\le$ $t_{high}^{2}$ and 
$t_{low}^{3}$ $<$ $(t_{D}-t_{C})$ $\le$ $t_{high}^{3}$.

In general, an $N$-node serial episode is associated with, $N-1$ inter-event constraints 
of the form $(t_{low}^{i},t_{high}^{i}]$. 
Episode discovery under such constraints involves discovery of frequent serial 
episodes along with discovery of the most appropriate inter-event constraint for 
every pair of nodes. In this subsection we present an algorithm for this where the 
user provides a set of non-overlapped intervals to serve as candidates for inter-event 
time constraints. An important special case is one where the same interval is to be 
used for all inter-event constraints and our general algorithm can easily be 
specialized for this case. 


\subsubsection{Candidate generation scheme}

The candidate generation schemes in \cite{Mannila1997,Srivats2005} require that the 
frequency of an episode is less that or equal to that of all its subepisodes. This is 
not true when we have inter-event time constraints. 
For example, if episode $(A^{\underrightarrow{(0,5]}}$  
$B^{\underrightarrow{(5,10]}}$ $C)$ is frequent, the sub-episodes 
$(A^{\underrightarrow{(0,5]}}$ $B)$ and $(B^{\underrightarrow{(5,10]}}$ $C)$ would be 
frequent, but for the subepisode 
$(A^{\underrightarrow{(?,?]}}$ $C)$ the inter-event constraint 
is not intuitive. Hence, we use a different candidate generation scheme  
here.

\begin{figure}[!htb]
\centering
\input{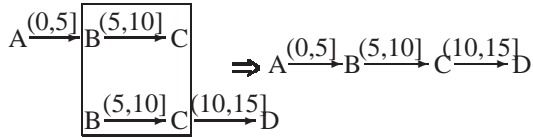}
\caption{\label{fig:Visualization-of-cand-gen-DISCOVERY}Visualization of
Candidate generation for serial episodes with inter-event constraints}
\end{figure}

The candidate episodes in this case are generated as follows. Let
$\alpha$ and $\beta$ be two $k$-node frequent episodes such that
by dropping the first node of $\alpha$ and the last node of $\beta$,
we get exactly the same $(k-1)$-node episode. 
A candidate episode $\gamma$ is generated by copying
the $k$-event types and $(k-1)$-intervals of $\alpha$ into $\gamma$
and then copying the last event type of $\beta$ into the $(k+1)^{th}$
event type of $\gamma$ and the last interval of $\beta$ to the $k^{th}$
interval of $\gamma$. Fig.~\ref{fig:Visualization-of-cand-gen-DISCOVERY} shows 
the candidate generation process graphically.

\subsubsection{Counting episodes with generalized inter-event time constraint}

We first explain the need for a new
algorithm to count occurrences of serial episode under inter-event time 
constraints. 
 Consider the event sequence
\begin{equation}
\langle(A,1),(A,2),(B,4),(A,5),(C,10),(B,12),(C,13),(D,17)\rangle.
\label{eq:interval-discovery}
\end{equation}
Let the serial episode under consideration be $(A^{\underrightarrow{(0,5]}}$ 
$B^{\underrightarrow{(5,10]}}$ $C^{\underrightarrow{(0,5]}}$ $D)$.
All the current algorithms for counting  occurrences of serial episodes 
either look at left most occurrence of episode or inner most occurrence of 
episode (See \cite{Srivats2006} for details). 
In the given event sequence, the left most occurrence 
is $\langle(A,1),$ $(B,4),$ $(C,10),$ $(D,17)\rangle$ and the inner most 
occurrence  is $\langle(A,5),$ $(B,12),$ $(C,13),$ 
$(D,17)\rangle$, where as only the occurrence $\langle(A,2),$ $(B,4),$ $(C,13),$ 
$(D,17)\rangle$  satisfies the inter-event constraints.

The counting algorithm is listed as Algorithm 2 in the Appendix. 
It uses $waits$ lists indexed by event types 
 and a linked list of $node$ structures 
for each episode as the basic data-structures.  
The entries in the $waits$ lists are $node$s. For each episode we have a 
doubly linked list of $node$ structures with a $node$ corresponding to each of the event 
types and arranged in the same order as that of the episode. 
The $node$ structure has a $tlist$ field that stores the times of 
occurrence of the event-type represented by its corresponding $node$. 
Other field in the $node$ structure is $visited$, 
which is a boolean field that indicates whether the event type is seen atleast once.

\begin{figure}[!htb]
\centering
\epsfig{file=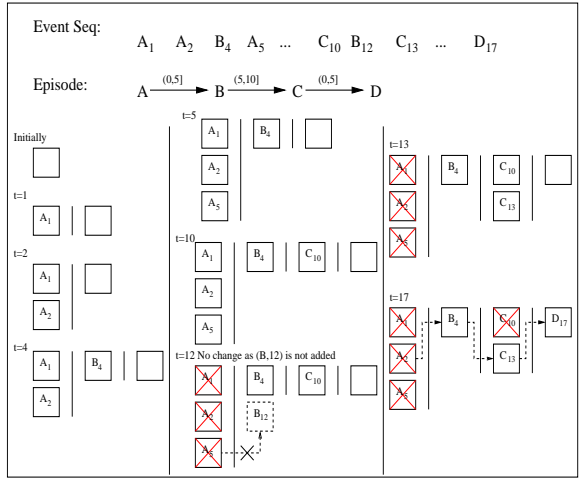, height=2.5in, width=3in}
\caption{\label{fig:Visualization-of-interval-DISCOVERY}Visualization of
Algorithm~\ref{alg:count-serial-INTERVAL-discovery}}
\end{figure}

On seeing an event type $E_{i}$, the algorithm iterates over list $waits(E_{i})$ 
and updates each $node$ in the list. We explain the procedure for updating the $node$s 
by considering the the example sequence given in 
(\ref{eq:interval-discovery}) and the episode 
$\alpha=(A^{\underrightarrow{(0,5]}}$ $B^{\underrightarrow{(5,10]}}$ 
$C^{\underrightarrow{(0,5]}}$ $D)$. Working of the algorithm in this example is 
illustrated in Fig.~\ref{fig:Visualization-of-interval-DISCOVERY}.

The $waits$ lists are initialized by adding the $node$s corresponding
to first event type of each episode in the set of candidates to the corresponding 
$waits(.)$ list. In the example, let the $node$
tracking event type $A$ be denoted by $node_{A}$, and so on. 
Initially $waits(A)$ contains $node_{A}$.
The boxes in Fig.~\ref{fig:Visualization-of-interval-DISCOVERY}
represent an entry in the $tlist$ of a $node$. An empty box is one
that is waiting for the first occurrence of an event type. On seeing
$(A,1)$, it is added to $tlist$ of $node_{A}$, and $node_{B}$
is added to $waits(B)$. At any time, \emph{the $node$ structures
are waiting for all event types that have been already seen and the
next unseen event type}.

Sometime later, at $t=4$, the first occurrence of a $B$ is seen. The $tlist$
of $node_{A}$ is traversed to find atleast one occurrence \emph{of
$A$,} such that $t_{B}-t_{A}\in(0,5]$. Both $(A,1)$ and $(A,2)$
satisfy this and hence, $(B,4)$ is accepted
into the $node_{B}.tlist$. The rule for accepting an occurrence of
an event type (which is not the first event type of the episode) is
that \emph{there must be atleast one occurrence of the previous event
type} (in this example $A$) \emph{which can be paired with the occurrence
of the current event type} (in this example $B$) \emph{without violating
the inter-event constraint}. Note that this check is not necessary for 
the first event of the episode. 
After seeing the first occurrence of $B$, $node_{C}$ is added to $waits(C)$.
Using the above rules the algorithms accepts $(A,5)$, $(C,10)$ into
the corresponding $tlist$s. At $t=12$, for $(B,12)$ none of the
entries in $node_{A}.tlist$ satisfy the inter-event constraint for
the pair $A\rightarrow B$. Hence $(B,12)$ is not added to the $tlist$
of $node_{B}$. Rest of the steps of the algorithm are illustrated in the figure.

If an occurrence of event type is added to $node.tlist$, it is because there 
exist events for each event type from the first to the event type corresponding 
to the $node$, which satisfy the respective inter-event time constraints.
An occurrence of episode is complete when an occurrence of the last
event type can be added to the $tlist$ of the last $node$ structure
tracking the episode.

The $tlist$ entries shown crossed out in the figure are the ones
that can be deallocated from the memory. 
In the example, at $t=12$, when the algorithm
tries to insert $(B,12)$ into $node_{B}.tlist$, the list of $tlist$
entries for occurrences of $A$'s is traversed. $(A,1)$ with inter-event
constraint $(0,5]$ can no longer be paired with a $B$ since the
inter-event time duration for any incoming event exceeds $5$, hence
$(A,1)$ can be safely removed from the $node_{A}.tlist$. This holds
for $(A,2)$ and $(A,5)$ as well. In this way the algorithm frees
memory wherever possible without additional processing burden.

Many times, in addition to counting frequencies, we may want to be able to track 
all the occurrances of episodes that were counted. For this, 
 we need to store sufficient back references 
in the data. This adds some memory overhead, 
but tracking may be useful in visualizing the discovered episodes.

\subsection{\label{sec:parallel-with-expiry}Parallel episodes with
expiry}

Since parallel episodes do not need any order on the events, it is reletively 
simpler to count their occurrances. 
We specialized the parallel episode discovery algorithm presented in \cite{Srivats2006}
to handle expiry time constraint. That is, we count the 
number of non-overlapped occurrences of a set of parallel episodes in which all the 
constituting events occur within time $T_{x}$ of each other. Since the modifications 
needed are simple, due to space limitations we do not provide the  
details here. 

\section{Simulation Results}
\label{sec:res}

In this section we present some results obtained using synthetic data as well as 
some real neuronal data. We also discuss the issue of statistical significance of 
discovered episodes. 

We used a simulation model to generate data that would resemble actual multi-neuronal 
recordings.  Spike train of each neuron is modeled as an
inhomogeneous poisson process. Neurons in the network are randomly 
interconnected.  Each connection is assigned a weight. 
For the random interconnections, the weight attached to each synapse is set 
 using a uniform distribution over $[-c, \; c]$ where $c$ is chosen to
be relatively small. When we want to embed any specific pattern, then, we set 
the weights of the required connections between neurons to a higher value.

The spike trains of each neuron is simulated as a rate varying poisson process.
The spiking rates of  neurons are 
updated every $\Delta T$  using the following.
\begin{equation}
\lambda_j(k) = \frac{\lambda_{\mbox{max}}}{1 + \exp{(-I_j(k) + d)}}
\label{eq:lambda-update}
\end{equation}
where $\lambda_j(k)$ is the firing rate of $j^{th}$ neuron at time $k \Delta T$ 
and $I_j(k)$ is its total input at that time. $I_j(k) = \sum O_i(k) w_{ij}$ 
where $O_i(k)$ is the output of $i^{th}$ neuron  
 and $w_{ij}$ is the weight of synapse from $i^{th}$ to $j^{th}$ neuron.
$O_i(k)$ is taken to be the number of spikes by the $i^{th}$ neuron in the 
interval $(\;(k-h) \Delta T, \ (k-h-1) \Delta T]$ where $h$ is a 
small integer that represents the 
synaptic delay in units of $\Delta T$.   
In (\ref{eq:lambda-update}, $\lambda_{max}$ is the maximum firing rate 
 and $d$ determines the resting spiking 
rate (i.e. when input is zero). This is the quiescent firing rate 
(or the noise level) in the system. An absolute refractory period is also used. 
This is the short time after a spike in which the neuron cannot respond to 
another stimulus.

\subsection{Network patterns}

In this section, we demonstrate how we can obtain useful information
about the structure of the underlying network using combination of
serial and parallel episode discovery. Using the simulation model
described above, we can embed different types of network patterns.
Fig. \ref{fig:episode-structures} shows examples of types of inter connections 
we make to embed different patterns. For this in the simulator 
we make these required connections between neurons have high weights. 
(In addition there are also random interconnections among neurons). 
We discuss three examples in this section.

\subsubsection*{Example 1}
\begin{figure}[!htb]
\centering
\input{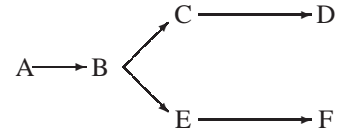}
\caption{\label{fig:pattern-1}Network pattern for Example 1}
\end{figure}

In a 26 neurons network (where each neuron corresponds to an alphabet)
we embed the pattern shown in Fig.\ref{fig:pattern-1}. The simulation
is run for 50 sec and approximately 25,000 spikes are generated. The
synaptic delay is set to be about 5 milli sec. We have chosen 
$\Delta T = 1$ milli sec and have taken refractory time also to be 
the same.

\begin{table}[!htb]
\centering
\begin{tabular}{|r|r|r|r|l|}
 \hline
Episode & Freq. & Time & Size & Patterns \\
expiry & Th. & (sec) & (No.) & Discovered \\
\hline
.0001 & .01 & .23 & 1(26) & \small{no episode of 2} \\
 & & & & \small{or more nodes} \\ \hline
.001 & .01 & .29 & 2(2) & \small{E C : 799; F D : 624} \\ \hline
.002 & .01 & .28 & 2(2) & \small{E C : 804; F D : 643} \\ \hline
.007 & .01 & .37 & 2(2) & \small{F E D C : 615} \\ \hline
\end{tabular}  
\caption{\label{tab:pattern-1-parallel}Parallel episodes for Example 1}
\end{table}

\begin{table}[!htb]
\centering
\begin{tabular}{|r|r|r|r|l|}
 \hline
Inter-event & Freq. & Time & Size & Patterns \\
interval & Th. & (sec) & (No.) & Discovered \\
 \hline
.000-.001 & .01 & .29 & 2(4) & \small{C E : 410; E C : 400} \\
 &  &  &  & \small{D F : 329; F D : 303} \\ \hline
.000-.002 & .01 & .31 & 2(4) & \small{C E : 422; E C : 408} \\
 &  &  &  & \small{D F : 348; F D : 323} \\ \hline
.002-.004 & .01 & .26 & 1(26) & \small{no 2 or more} \\ 
	&	&	&	& \small{node episodes} \\ \hline
.004-.006 & .01 & .29 & 4(4) & \small{A B C D : 597} \\
 &  &  &  & \small{A B E F : 589} \\
 &  &  &  & \small{A B E D : 530} \\
 &  &  &  & \small{A B C F : 530} \\ \hline
\end{tabular}  
\caption{\label{tab:pattern-1-serial}Serial episodes Example 1}
\end{table}

The sequence is then mined for frequent parallel episodes with
different expiry times. The results are given in Table
\ref{tab:pattern-1-parallel}. The table shows the expiry time used, 
the frequency threshold, time taken by the algorithm on a Intel dual core 
PC running at 1.6 GHz, the size of the largest frequent episode discovered and 
the number of episodes of this size along with the actual episodes. We follow 
the same structure for all the tables. The frequency threshold is expressed as
a fraction of the entire data length. A threshold of 0.01 over a data
length of 25,000 spike events requires an episode to occur atleast 250
times before it is declared as frequent. From Table
\ref{tab:pattern-1-parallel} it can be seen that $(CE)$ and $(DF)$
turn out to be the only frequent parallel episodes if the 
expiry time is 1 to 2 milli sec. If the expiry time is too small, we get 
no frequent episodes (at this threshold). On the other hand, if we increase 
the expiry time to be 7 milli sec which is greater than a synaptic delay, 
then even $(FEDC)$ turns out to be a parallel episode. This shows that by using 
appropriate expiry time, parallel episodes discovered 
capture synchronous firing patterns. 

The results of serial episode discovery are shown in Table
\ref{tab:pattern-1-serial}. With an inter-event constraint of 4-6
milli sec, we discover all paths in the network (Fig.
\ref{fig:pattern-1}). When we prescribe that inter-event time be less 
than 2 milli sec (when synaptic delay is 5 milli sec), we get nodes in the 
same level as our serial episodes. If we use intervals of 2-4 milli sec, we 
get no episodes because synchronous firings mostly occur much closer 
and firings related by a synapse have a delay of 5 milli sec. Thus, using 
inter-event time constraints, we can get fair amount of information of the 
underlying connection structure.  
It may seem surprising that we also discover
$A\rightarrow B$ $\rightarrow C$ $\rightarrow F$ and $A\rightarrow B$
$\rightarrow E$ $\rightarrow D$ when we use 4--6 milli sec constraint. 
This is because, the network structure is 
such that $D$ and $F$ fire about one synaptic delay time 
after the firing of $C$ and $E$. Thus, the serial episodes give the sequential 
structure in the firings which could, of course, be generated by different 
interconnections. The frequent episodes discovered 
provide a handle to unearthing the hierarchy seen in the data (i.e.
which events co-occur and which ones follow one another).

\subsubsection*{Example 2}
In this example we consider the network connectivity pattern as shown in
Fig.~\ref{fig:episode-structures}(c). As stated earlier, this is an example 
of a  
Synfire chain. We use the same parameters in the simulator as in Example 1
 and generate spike trains data using this connectivity pattern.  
Table~\ref{tab:pattern-3-parallel} shows the parallel episodes discovered  
 and Table~\ref{tab:pattern-3-serial} shows the serial episodes discovered 
with different inter-event constraints. From the tables, it is easily seen 
that parallel episodes with expiry time of 1 milli sec and 
serial episodes with inter-event time constraint of about one synaptic delay, 
together give good information about underlying network structure. In this example, 
we illustrate how our algorithms can discover synfire chain patterns. We first 
discover all parallel episodes with expiry time 1 milli sec. Then for  
each frequent parallel episode,  
we replace each of its occurrences in the data stream by a new event with event type 
being the name of the parallel episode. This new event is put in with a time 
of occurrence which is the mean time in the episode occurrence. We then discover 
all serial episodes with different inter-event time constraints. The 
results obtained with this method are shown in Table~\ref{tab:pattern-3-synfire}. 
As can be seen, 
the only pattern we discover is the underlying synfire chain. This example shows 
that by proper combination of parallel and serial episodes, we can obtain fairly 
rich pattern structures which are of interest in neuronal spike train analysis. 

\begin{table}[!htb]
\centering
\begin{tabular}{|r|r|r|r|l|}
 \hline
Episode & Freq. & Time & Size & Patterns \\
expiry & Th. & (sec) & (No.) & Discovered \\
\hline
.001 & .01 & .15 & 4(1) & \small{L K : 307} \\
 &  &  &  & \small{C B D : 293} \\
 &  &  &  & \small{H G F I : 268} \\
 &  &  &  & \small{rest are} \\
 &  &  &  & \small{sub-episodes} \\ \hline
\end{tabular}  
\caption{\label{tab:pattern-3-parallel}Parallel episodes for Example 2}
\end{table}

\begin{table}[!htb]
\centering
\begin{tabular}{|r|r|r|r|l|}
 \hline
Inter-event & Freq. & Time & Size & Patterns \\
interval & Th. & (sec) & (No.) & Discovered \\
\hline
.002-.004 & .01 & .157 & 1(26) & \small{no episodes of 2} \\
 & & & & \small{or more nodes} \\ \hline
.004-.006 & .01 & .469 & 6(24) & \small{A D E H J K : 195} \\
 &  &  &  & \small{A D E I J K : 194} \\
 &  &  &  & \small{A D E H J L : 193} \\
 &  &  &  & \small{A C E H J K : 192} \\ \hline
.006-.008 & .01 & .156 & 1(26) & no episodes of 2 \\
 & & & & or more nodes \\ \hline
\end{tabular}  
\caption{\label{tab:pattern-3-serial}Serial episodes for Example 2}
\end{table}

\begin{table}[!htb]
\centering
\begin{tabular}{|r|r|r|r|l|}
 \hline
Inter-event & Freq. & Time & Size & Patterns \\
interval & Th. & (sec) & (No.) & Discovered \\
\hline
.002-.004 & .01 & .11 & 1(20) & \small{no episodes of} \\
 & & & & \small{2 or more nodes} \\ \hline
.004-.006 & .01 & .14 & 6(1) & \small{A [C B D] E} \\
 & & & & \small{[H G F I] J [L K] : 137} \\ \hline
.006-.008 & .01 & .12 & 1(20) & \small{no episodes of} \\
 & & & & \small{2 or more nodes} \\ \hline
\end{tabular}  
\caption{\label{tab:pattern-3-synfire}Synfire episodes for Example 2}
\end{table}

\subsubsection*{Example 3}
\begin{figure}[!htb]
\centering
\input{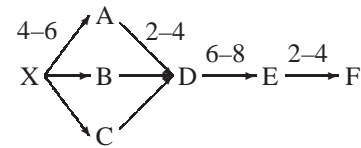}
\caption{\label{fig:pattern-2}Network Pattern for Example 3}
\end{figure}

In this example, we choose a network pattern where different pairs of 
interconnected neurons can have different synaptic delays 
and we demonstrate the ability of our algorithm to 
automatically discover appropriate inter-event intervals. 
The pattern is shown in Fig.
\ref{fig:pattern-2}, where we have different synaptic delays 
as indicated on the figure.

\begin{table}[!htb]
\centering
\begin{tabular}{|r|r|r|r|l|}
 \hline
Episode & Freq. & Time & Size & Patterns \\
expiry & Th. & (sec) & (No.) & Discovered \\
\hline
.001 & .01 & .28 & 3(1) & \small{A B C : 614} \\ \hline
.002 & .01 & .25 & 3(1) & \small{A B C : 617} \\ \hline
.004 & .01 & .28 & 4(1) & \small{A B C D : 537} \\ \hline
.006 & .01 & .32 & 4(2) & \small{X A B C : 602} \\
 &  &  &  & A B C D : 542 \\ \hline
\end{tabular}  
\caption{Parallel episodes for Example 3}
\label{tab:pattern-2-parallel}
\end{table}

\begin{table}[!htb]
\centering
\begin{tabular}{|r|r|r|r|l|}
 \hline
Inter-event & Freq. & Time & Size & Patterns \\
interval & Th. & (sec) & (No.) & Discovered \\
\hline
.000-.002 & .01 & .32 & 2(6) & \small{A C : 385; B A : 376} \\
 &  &  &  & \small{B C : 373; A B : 372} \\
 &  &  &  & \small{C A : 361; C B : 355} \\ \hline
.002-.004 & .01 & .37 & 2(4) & \small{E F : 783; A D : 656} \\
 &  &  &  & \small{C D : 651; B D : 646} \\ \hline
.004-.006 & .01 & .28 & 2(3) & \small{X A : 790; X B : 774} \\
 &  &  &  & \small{X C : 769} \\ \hline
.006-.008 & .01 & .29 & 2(2) & \small{D E : 720; X D : 454} \\ \hline
\end{tabular}  
\caption{Serial Episodes for Example 3}
\label{tab:pattern-2-serial}
\end{table}

\begin{table}[!htb]
\centering
\begin{tabular}{|r|r|r|r|l|}
 \hline
Inter-event & Freq. & Time & Size & Patterns \\
interval & Th. & (sec) & (No.) & Discovered \\
\hline
\{.000-.002, & .01 & 1.37 & 5(1) & \small{$X^{\underrightarrow{.004-.006}}$} \\
.002-.004, &  &  &  & \small{$[A B C]^{\underrightarrow{.002-.004}}$}  \\
.004-.006, &  &  &  &  \small{$D^{\underrightarrow{.006-.008}}$} \\
.006-.008, &  &  &  & \small{$E^{\underrightarrow{.002-.004}}F$}\\
.008-.010\} &  &  &  & \small{ : 372} \\ \hline
\end{tabular}  
\caption{\label{tab:pattern-2-synfire}Synfire episodes for Example 3}
\end{table}

The results for parallel episode discovery (see Table
\ref{tab:pattern-2-parallel}) show that $(ABC)$ is the group of
neurons that co-spike together. The serial episode discovery results
are given in Table \ref{tab:pattern-2-serial}. As can be seen from 
the table, with different 
 pre-specified inter-event time constraints we can discover only 
different parts of the underlying network graph because no single 
inter-event constraint captures the full pattern. 

As in Example 2, we replace occurrences of parallel episode with a 
new event in the data stream. We then run Algorithm~\ref{alg:count-serial-INTERVAL-discovery} to 
discover serial episodes along with inter-event constraints, given a set of 
possible inter-event intervals. The results obtained are shown in 
Table \ref{tab:pattern-2-synfire}. As can be seen from the table, 
the algorithm is very effective in unearthing the underlying 
network pattern.

\subsection{Significance of discovered patterns}

The examples above show that if we generate spike data using special embedded 
patterns in it then our algorithms can detect them. However, this does not answer the 
question: if the algorithm detects some frequent episodes what confidence do we have 
that they correspond to some patterns. Here, we try to answer this question by showing 
that it is unlikely to have long frequent episodes if the data generation model does 
not have any specific biases. We generate such random data as follows. We use the 
same simulator but with only random interconnection weights and no specially introduced 
strong connections. We generated ten sets of random interconnection weights and for 
each set we generated ten sets of data (25 000 spikes) 
by running the simulator with those weights. Thus 
we have 100 data sets in which while neurons still fire with input dependent firing 
rates, there are no special causative connections. Apart from this we generated another 
50 data sets where the firing rates of neurons are chosen randomly at each $\Delta T$. 
We then discover serial episodes  of size upto 10 with a 
frequency threshold of zero so that we get frequencies for all episodes. 
Table~\ref{tab:noise-serial} shows {\em maximum} frequency (averaged over the 150 
data sets) versus size of episodes that we obtained. 
We have also generated 20 data sets in which a 
long ordered chain is embedded. The table also shows the {\em minimum} frequency 
(averaged over 
the 20 data sets) versus size for episodes which are subepisodes of the embedded chain. 
From the table it can be seen that even for size 2, the maximum frequency of an episode 
in the random data is very small. From size 3 onwards, all episodes have frequency less 
than 10. On the other hand, when the data contains patterns, 
even the minimum observed frequencies of that size 
episodes are about two orders of magnitude larger. This provides sufficient statistical 
justification that it is higly unlikely to have long episodes with appreciable frequencies 
if the data source does not have the necessary bias. 

\begin{table}[!htb]
\centering
\begin{tabular}{|r|r|r|}
\hline
\multicolumn{1}{|c|}{Size} & \multicolumn{2}{c|}{26 event types}\\ \cline{2-3}
	&  Noise sequence & Sequences with patterns \\ \cline{2-3}
	& Avg. Max. Episode & Avg. Min. Sub-Episode \\
	& Frequency & Frequency \\ \hline
1-Node &    1050.57 &     967.80 \\ \hline
2-Node &      61.47 &     845.65 \\ \hline
3-Node &       8.51 &     734.55 \\ \hline
4-Node &       3.31 &     647.30 \\ \hline
5-Node &       2.03 &     576.06 \\ \hline
6-Node &       1.25 &     515.88 \\ \hline
7-Node &       .13 &     466.33 \\ \hline
8-Node &       .12 &     423.58 \\ \hline
9-Node &       .12 &     385.25 \\ \hline
10-Node &      .12 &     353.88 \\ \hline
	& Sample size = 150 & Sample size = 20 \\ \hline
\end{tabular}
\caption{\label{tab:noise-serial} Serial Episode frequencies in random and patterned data}
\end{table}

\subsection{\label{sec:Analysis-of-multi-neuron}Analysis of multi-neuron data}

\begin{figure}[!htb]
\centering
\subfigure[Frequent parallel episodes of size 10 satisfying expiry constraint = 10 time units]{\epsfig{file=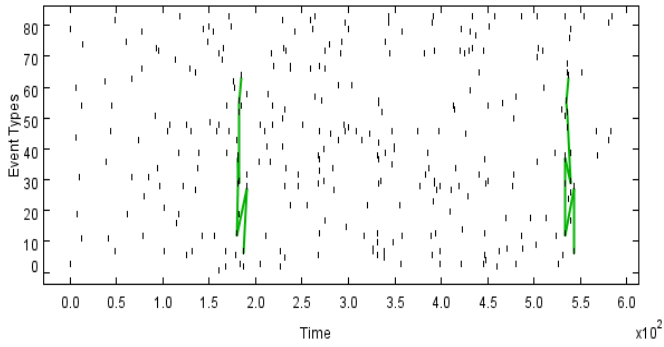,width=3.5in,height=2in}}
\subfigure[Frequent serial episodes of size 4 satisfying inter-event constraint = 10 time units]{\epsfig{file=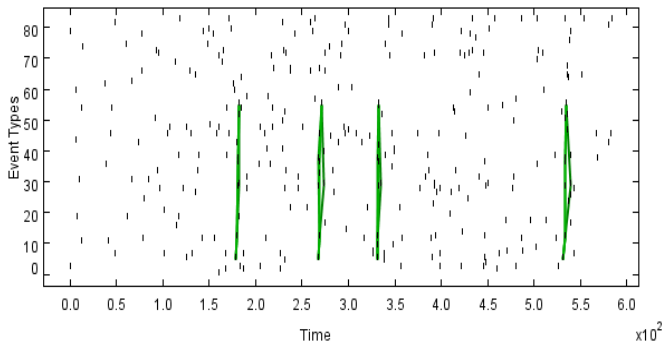,width=3.5in,height=2in}}
\caption{\label{fig:our-results}Frequent episodes discovered using our algorithms on real data.}
\end{figure}

In this section we describe results obtained on calcium imaging data reported 
in \cite{Ikegaya2004}. (We are grateful to Dr. Rafeal Yuste for sharing this 
Calcium Imaging data with us). In \cite{Ikegaya2004}, Ikegaya et. al. analyzed how neural 
activity propagates through cortical networks. They found precise repetitions
of spontaneous patterns. These patterns repeated after minutes maintaining
millisecond accuracy. In Fig.~3A of \cite{Ikegaya2004}, such patterns 
are shown in raster plots by connecting the spikes that are part of
an occurrence.

In Fig.~\ref{fig:our-results}, we show results obtained on the same
calcium imaging data set using frequent episode discovery algorithms.
Fig.~\ref{fig:our-results}~(a) shows two occurrences of a 10-node
parallel episode discovered 
with expiry time constraint $T_{X}=10$ time units. Fig.~\ref{fig:our-results}~(b)
shows four occurrences of two 4-node serial episode discovered 
 with inter-event constraint
of 0 to 10 time units. It is seen that the results obtained
using frequent episode discovery match with those presented in \cite{Ikegaya2004}.
Also, the time needed by our algorithm is much smaller because in 
\cite{Ikegaya2004}, they use a counting technique that cannot control the 
combinatorial explosion. 
This result brings out the utility of our data mining 
 technique in terms of both effectiveness and efficiency.

\section{Conclusion}
\label{sec:conc}

Frequent episode discovery is
a very efficient temporal data mining technique. In this paper we 
have presented some new algorithms for frequent episode discovery 
under expiry time and inter-event time constraints. 
The temporal constraints are motivated by the problem of analyzing 
multi-neuron spiking data. We have discussed the kind of patterns 
that neurobiologists look for in such data and have shown that our algorithms 
are very effective in unearthing the underlying connectivity structure from 
spike data. In this context, our temporal constraints are very useful in 
focusing the search for patterns and tackling combinatorial explosion. 
Also, we can readily relate these constraints to relevant biological parameters. 

One of the main motivations for this paper is to introduce the problem 
multi-neuron spike data analysis to data mining community. This is a 
challenging problem of analyzing large data sets to find underlying 
patterns, though no data mining techniques have so far been used for this. 
One can think of this problem as one of learning network connectivity 
pattern given only node-level dynamic data. Such a problem would be relevant 
in many other application areas as well. For example, analyzing abnormal 
behavior of communication networks, finding hidden causative chains in 
complex manufacturing processes controlled by distributed controllers, etc. 
We hope our results presented here would contribute towards developing of 
data mining techniques relevant in such applications. 

\bibliographystyle{abbrv}
\bibliography{frequent_episodes}

\appendix
\section{Pseudo-code listing}

\begin{algorithm}[!htb]
\caption{\label{alg:count-serial-INTERVAL-discovery}Counting serial
episodes with inter-event time constraints}
\begin{algorithmic}[1]

\REQUIRE Set $C$ of $N$-node episodes, event
streams $\langle(E_{i},t_{i})\rangle$, frequency
threshold $\lambda_{min}\in[0,1]$
\ENSURE The set $F$ of frequent episodes in $C$
\FORALL{event types $A$}
	\STATE Initialize $waits(A)=\phi$
\ENDFOR
\FORALL{$\alpha\in C$}
	\STATE Set $prev=\phi$
	\FOR{$i=1$ to $N$} 
		\STATE Create $node$ with $node.visited=false$; $node.episode=\alpha$;
		$node.index=i$; $node.prev=prev$; $node.next=\phi$
		\IF{$i=1$}
			\STATE Add $node$ to $waits(\alpha[1])$ 
		\ENDIF
		\IF{$prev\ne\phi$}
			\STATE $prev.next=node$
		\ENDIF
	\ENDFOR
\ENDFOR
\FOR{$i=1$ to $n$}
	\FORALL{$node\in waits(E_{i})$}
		\STATE Set $accepted=false$; $\alpha=node.episode$; $j=node.index$; $tlist=node.tlist$
		\IF{$j<N$}
			\FORALL{$tval\in tlist$}
				\IF{$(t_{i}-tval.init)>\alpha.t_{high}[j]$}
					\STATE Remove $tval$ from $tlist$
				\ENDIF
			\ENDFOR
		\ENDIF
		\IF{$j=1$}
			\STATE Update $accepted=true$; $tval.init=t_{i}$
			\STATE Add $tval$ to $tlist$
			\IF{$node.visited=false$}
				\STATE Update $node.visited=true$
				\STATE Add $node.next$ to $waits(\alpha[j+1])$ 
			\ENDIF
		\ELSE
			\FORALL{$prev\_tval\in node.prev.tlist$}
				\IF{$t_{i}-prev\_tval\in(\alpha.t_{low}[j-1],\alpha.t_{high}[j-1]]$}
					\STATE Update $accepted=true$; $tval.init=t_{i}$
					\STATE Add $tval$ to $tlist$
					\IF{$node.visited=false$}
						\STATE Update $node.visited=true$
						\IF{$node.index\le N-1$}
							\STATE Add $node.next$ to $waits(\alpha[j+1])$
						\ENDIF
					\ENDIF
				\ELSE
					\IF{$t_{i}-prev\_tval>\alpha.t_{high}[j-1]$}
						\STATE Remove $prev\_tval$ from $node.prev.tlist$
					\ENDIF
				\ENDIF
			\ENDFOR
		\ENDIF
		\IF{$accepted=true$ and $node.index=N$}
			\STATE Update $\alpha.freq=\alpha.freq+1$
			\STATE Set $temp=node$
			\WHILE{$temp\ne\phi$}
				\STATE Update $temp.visited=false$
				\IF{$temp.index\ne1$}
					\STATE Remove $temp$ from $waits(\alpha[temp.index])$
				\ENDIF
				\STATE Update $temp=temp.next$
			\ENDWHILE
		\ENDIF
	\ENDFOR
\ENDFOR
\STATE Output $F=\{\alpha\in C$ such that $\alpha.freq\ge n\lambda_{min}\}$
\end{algorithmic}
\end{algorithm}

\end{document}